\documentclass[aps,pra,floatfix,12]{revtex4-2}
\usepackage{amsfonts}
\usepackage{amsmath}
\usepackage{amsthm}
\usepackage{amssymb}
\usepackage{graphicx}
\usepackage{appendix}
\usepackage{hyperref}
\usepackage{textcomp}
\usepackage{multirow}
\usepackage{color}
\usepackage{bm}
\usepackage{braket}
\usepackage{subfigure}
\usepackage{setspace}

\begin{document}

\title{Interactions between fractional solitons in bimodal fiber cavities}
\author{Tandin Zangmo$^{1}$, Thawatchai Mayteevarunyoo$^{1}$, and Boris A.
Malomed$^{2,3}$}
\address{$^{1}$Department of Electrical and Computer Engineering, Faculty of
	Engineering, Naresuan University, Phitsanulok 65000, Thailand\\
	$^{2}$Department of Physical Electronics, School of Electrical Engineering,
	Faculty of Engineering, and the Center for Light-Matter University, Tel Aviv
	University, Tel Aviv, Israel\\
	$^{3}$Instituto de Alta Investigaci\'{o}n, Universidad de Tarapac\'{a},
	Casilla 7D, Arica, Chile}

\begin{abstract}
We introduce a system of fractional nonlinear Schr\"{o}dinger equations
(FNLSEs) which model the copropagation of optical waves carried by different
wavelengths or mutually orthogonal circular polarizations in fiber-laser
cavities with the effective fractional group-velocity dispersion (FGVD),
which were recently made available to the experiment. In the FNLSE system,
the FGVD terms are represented by the Riesz derivatives, with the respective
Levy index (LI). The FNLSEs, which include the nonlinear
self-phase-modulation (SPM) nonlinearity, are coupled by the cross-phase
modulation (XPM) terms, and separated by a group-velocity (GV) mismatch (%
\textit{rapidity}). By means of systematic simulations, we analyze
collisions and bound states of solitons in the XPM-coupled system, varying
the LI and GV mismatch. Outcomes of collisions between the solitons include
rebound, conversion of the colliding single-component solitons into a pair
of two-component ones, merger of the solitons into a breather, their mutual
passage leading to excitation of intrinsic vibrations, and the elastic
interaction. Families of stable two-component soliton bound states are
constructed too, featuring a rapidity which is intermediate between those of
the two components.

\smallskip \noindent \textbf{Keywords}: Fractional nonlinear Schr\"{o}dinger
equations; Riesz fractional derivative; group-velocity dispersion; Kerr
nonlinearity; wavelength-division multiplexing; inelastic collisions of
solitons; two-component solitons
\end{abstract}

\maketitle

{\Large Dedicated to the memory of Professor David J. Kaup (1939 -- 2022)}

\section{Introduction and the model}

One of many fundamental contributions of David Kaup to mathematical and
theoretical physics is the analysis of interactions (alias \textit{crosstalk}%
) between solitons carried by different channels in
wavelength-division-multiplexed (WDM) fiber-optic telecommunication lines
\cite{Kaup}. The significance of this problem for applications is well known
\cite{WDM1,WDM2,Hasegawa}, and it is also recognized as a subject of
interest to the theory of nonlinear-wave propagation in optical systems in
general \cite{Biondini,Peleg,Shtaif,Turitsyn}.

Recently, the traditional models of the linear and nonlinear propagation of
signals in dispersive optical fibers were extended to, and experimentally
realized in, fiber cavities with \textit{fractional} group-velocity
dispersion (FGVD) \cite{Shilong,Shilong-nonlin}. In fact, these works have
reported the first experimental realization of the wave propagation in
fractional media of any physical nature. The present work, as a contribution
to the special issue of Studies in Applied Mathematics dedicated to the
memory of David Kaup, aims to extend the analysis of the interaction between
solitons in XPM (cross-phase-modulation)-coupled nonlinear channels with
FGVD (instead of the ordinary non-fractional GVD). This topic is a basic
element of WDM implemented in the fiber cavity (in fact, it is a fiber laser
with compensated losses and gain) featuring the effective FGVD. In addition
to its significance to the current studies in nonlinear optics, this
problem, being closely related to the specific definition of the
pseudodifferential operator which represents the fractional derivative (see
Eq. (\ref{Riesz derivative}) below) is directly relevant to other topics
covered by Studies in Applied Mathematics.

While fractional derivatives were known for a long time as a formal
generalization of the classical calculus \cite{Abel}-\cite{Uchaikin}, they
were introduced in physics by Laskin in the form of fractional quantum
mechanics \cite{Lask1,Lask2} (see also Ref. \cite{GuoXu}), which is based on
the fractional Schr\"{o}dinger equation (FSE) for the wave function of
particles whose classical motion is performed by random \textit{L\'{e}vy
flights}. The latter means that the mean distance of a stochastically moving
particle from its initial position, $x=0$, grows with time as
\begin{equation}
|x|\sim t^{1/\alpha },  \label{|x|}
\end{equation}%
where the \textit{L\'{e}vy index} (LI) $\alpha $ takes values%
\begin{equation}
0<\alpha \leq 2  \label{LI}
\end{equation}%
\cite{Mandelbrot}. In the scaled form, the corresponding FSE for the
particle's wave function $\Psi (x,t)$ is written as
\begin{equation}
i\frac{\partial \Psi }{\partial t}=\frac{1}{2}\left( -\frac{\partial ^{2}}{%
\partial x^{2}}\right) ^{\alpha /2}\Psi +V(x)\Psi ,  \label{FSE}
\end{equation}%
where $V(x)$ is the usual external potential. The normal Schr\"{o}dinger
equation corresponds to $\alpha =2$, while in the case of $\alpha <2$ the
kinetic-energy operator in Eq. (\ref{FSE}) is replaced by the Riesz
fractional derivative \cite{Riesz}, which implies that the fractional
differentiation with LI $\alpha $ is performed by the multiplication of the
Fourier transform of the wave function, $\hat{\Psi}(p)$, by $|p|^{\alpha }$:
\begin{equation}
\left( -\frac{\partial ^{2}}{\partial x^{2}}\right) ^{\alpha /2}\Psi =\frac{1%
}{2\pi }\int_{-\infty }^{+\infty }dp|p|^{\alpha }\int_{-\infty }^{+\infty
}d\xi e^{ip(x-\xi )}\Psi (\xi ).  \label{Riesz derivative}
\end{equation}%
The fundamental property of Eq. (\ref{FSE}) is that a wave packet spreads
out in the free space so that its width $W$ grows with time according to the
same equation (\ref{|x|}), $W\sim t^{1/\alpha }$, i.e., faster than in the
usual case corresponding to $\alpha =2$. This conclusion is correlated with
the fact that the stochastic L\'{e}vy-flight motion, governed by Eq. (\ref%
{|x|}) with $\alpha <2$, is faster than the classical diffusion spread
corresponding to $\alpha =2$.

Implementations of the fractional quantum theory were sought for in
solid-state setups, including \textit{L\'{e}vy crystals} \cite{Levy crystal}
and exciton-polariton condensates in semiconductor microcavities \cite%
{Pinsker}. However, no experimental realization of these proposals has been
reported, as yet. An alternative realization of the FSE was proposed by
Longhi \cite{Longhi}, making use of the commonly known similarity of the
quantum-mechanical Schr\"{o}dinger equation to the parabolic propagation
equation for the local amplitude of the classical electromagnetic field
under the condition of the paraxial diffraction, replacing time $t$, as the
evolution variable, by the propagation distance in the waveguide, $z$ \cite%
{KA}. The so proposed optical $4f$ (four-focal-lengths) setup refers to a
cavity with two lenses and a phase mask placed in the midplane between the
lenses. The lenses carry out the direct and inverse Fourier transforms of
the light beam with respect to the transverse coordinate in the cavity. The
Fourier transform decomposes the beam into parallel-propagating spectral
components with transverse wavenumber $p$, the distance $d$ of each
component beam from the optical axis being a functions of $p$. The phase
mask is designed so that it imparts a phase shift $\Delta \phi (d(p))$ to
the passing component beam, emulating the phase variation $\Delta \phi _{
\mathrm{frac}}$ which should be produced by the desired fractional
diffraction in the Fourier space. According to Eqs. (\ref{FSE}) and (\ref%
{Riesz derivative}), $\Delta \phi _{\mathrm{frac}}=\exp \left( -i|p|^{\alpha
}Z\right) $, where $Z$ is the propagation distance corresponding to the
action of the fractional diffraction. Thus, one may emulate the effect of
the fractional diffraction, designing the phase mask with $\Delta \phi
(d(p))=\Delta \phi _{\mathrm{frac}}$. This setup provides a single-stage
(discrete) action of the effective fractional diffraction. One may then
approximate the continuous FSE (\ref{FSE}), considering many cycles of the
light circulation in the cavity, assuming that each cycle gives rise to a
small phase shift $\Delta \phi (d(p))$. The potential $V(x)$ in Eq. (\ref%
{FSE}) can be represented by a mirror inserted in the optical cavity \cite%
{Longhi}.

While the effective fractional diffraction, based on the setup outlined
above, has not yet been implemented experimentally, the realization of the
effective FGVD in fiber cavities was reported recently \cite{Shilong}. It
also relies on emulating the action of the FGVD in the Fourier domain by
dint of the differential phase shifts imparted to spatially separated
spectral-component beams. The phase shifts were imposed by an appropriately
designed phase mask, which was created as a computer-generated hologram.
Thus, the temporal optical signal propagating in the fiber was
Fourier-decomposed into spatially separated spectral components, that were
passed through the hologram and recombined back into the temporal domain.

The realization of the FSE as the propagation equation in optics suggests a
natural possibility to add effects of the material nonlinearity, thus
replacing FSEs by fractional nonlinear Schr\"{o}dinger equations (FNLSEs).
This possibility had drawn much interest in studies in the realm of
nonlinear optics at the theoretical level. Appropriate numerical methods for
the work combining the fractional derivatives, represented by the nonlocal
pseudodifferential operators (see Eq. (\ref{Riesz derivative})), and local
nonlinearities are available \cite{soliton1,soliton2}, which has made it
possible to predict many results for solitons in such systems with the cubic
\cite{Chen}-\cite{Strunin}, cubic-quintic \cite{Frac16,Frac11}, and
quadratic \cite{Liangwei} nonlinear terms. Other nonlinear states, such as
fractional discrete solitons \cite{Molina,electrical}, Airy waves \cite%
{soliton4} and domain walls \cite{Kumar}, were also predicted in the
framework of FNLSEs. Many theoretical results for fractional solitons, in
both one- and two-dimensional models, are summarized in recent reviews \cite%
{review,review2}.

The first experimental observation of nonlinear effects in a fiber cavity
combining the effective FGVD and Kerr (cubic) nonlinearity was reported very
recently \cite{Shilong-nonlin}. These are transitions (bifurcations) between
optical pulses with single- and multi-peak spectral shapes.

The present work addresses interactions between solitons in the two-channel
FGVD system with a group-velocity (GV) mismatch, $c$. The model, which is
based on the pair of XPM-coupled FNLSEs, is introduced in Section 2.
Systematic results for collisions between solitons are reported in Section
3, where they are summarized by means of a chart in the parameter plane of $%
c $ and LI, $\alpha $ (see Fig. \ref{fig13a} below). Section 4 produces
systematic results for families of stable two-soliton bound states, which
exist in spite of the presence of the group-velocity mismatch between the
channels. The paper is concluded by Section 5.

\section{The model}

In the scaled form, the system of FNLSEs for amplitudes $u$ and $v$ of the
copropagating optical fields in the XPM-coupled channels, carried by
different wavelengths, is written as
\begin{eqnarray}
i\frac{\partial u}{\partial z} &=&\frac{1}{2}\left( -\frac{\partial ^{2}}{%
\partial \tau ^{2}}\right) ^{\alpha /2}u+\left( |u|^{2}+2|v|^{2}\right) u,
\label{2u} \\
i\frac{\partial v}{\partial z} &=&-ic\frac{\partial v}{\partial \tau }+\frac{%
1}{2}\left( -\frac{\partial ^{2}}{\partial \tau ^{2}}\right) ^{\alpha
/2}v+\left( |v|^{2}+2|u|^{2}\right) v,  \label{2v}
\end{eqnarray}%
where $z$ and $\tau $ are. as usual, the propagation distance and reduced
time \cite{Agrawal}, while $\alpha $, as said above, is the LI which
determines the FGVD (cf. Eq. (\ref{Riesz derivative})). The sign of the FGVD
terms in Eqs. (\ref{2u}) and (\ref{2v}) corresponds to the anomalous
temporal dispersion, which is the case of major interest in fiber optics
\cite{Agrawal}, and real $c$ represents the mismatch in the inverse GV
between the channels, alias the relative \textit{rapidity}. The relative
coefficient $2$ in front of the XPM terms, in comparison to the SPM
(self-phase-modulation) ones, is another common feature of the light
propagation in fibers \cite{Agrawal}. Relevant values of LI are
\begin{equation}
1<\alpha \leq 2,  \label{LI2}
\end{equation}%
as in the remaining interval, $0<\alpha \leq 1$ (see Eq. (\ref{LI})), the
\textit{critical} or \textit{supercritical collapse} takes place (at $\alpha
=1$ or $\alpha <1$, respectively) \cite{Chen,Frac12,review,review2}, making
all solitons unstable.

Another realization of the same system of Eqs. (\ref{2u}) and (\ref{2v}) in
fiber optics, with the same ratio, $\mathrm{XPM:SPM}=2$, can be realized by
the copropagation of two mutually orthogonal circular polarizations of light
carried by the same wavelength. In that case, parameter $c$ represents the
GV birefringence, which may be imposed by the fiber's twist \cite{Agrawal}.

The system of Eqs. (\ref{2u}) and (\ref{2v}) conserves four dynamical
invariants, namely, the energies (norms) of the two components,%
\begin{equation}
E_{u}=\int_{-\infty }^{+\infty }\left\vert u(\tau )\right\vert ^{2}d\tau
,~E_{v}=\int_{-\infty }^{+\infty }\left\vert v(\tau )\right\vert ^{2}d\tau ,
\label{EE}
\end{equation}%
the total momentum,%
\begin{equation}
P=i\int_{-\infty }^{+\infty }\left( uu_{\tau }^{\ast }+vv_{\tau }^{\ast
}\right) d\tau ,  \label{P}
\end{equation}%
and the Hamiltonian%
\begin{gather}
H=-\frac{1}{2\pi }\int_{-\infty }^{+\infty }d\tau _{1}\int_{-\infty
}^{+\infty }d\tau _{2}\int_{-\infty }^{+\infty }d\omega |\omega |^{\alpha
}e^{i\omega (\tau _{1}-\tau _{2})}\left[ u^{\ast }\left( \tau _{1}\right)
u\left( \tau _{2}\right) +v^{\ast }\left( \tau _{1}\right) v\left( \tau
_{2}\right) \right] +ic\int_{-\infty }^{+\infty }d\tau v^{\ast }v_{\tau }
\notag \\
-\int_{-\infty }^{+\infty }d\tau \left( \frac{1}{2}|u|^{4}+\frac{1}{2}
|v|^{4}+2|u|^{2}|v|^{2}\right) .  \label{H2}
\end{gather}

In interval (\ref{LI2}), Eqs. (\ref{2u}) and (\ref{2v}) admit stable
single-component soliton solutions, \textit{viz}.,%
\begin{eqnarray}
u &=&\exp \left( ik_{u}z\right) U(\tau ),~v=0,  \label{v=0} \\
u &=&0,~v=\exp \left( ik_{v}z\right) V\left( \tilde{\tau}\right) ,~\tilde{%
\tau}\equiv \tau -cz+\tau _{0},  \label{u=0}
\end{eqnarray}%
where $k_{u}$ and $k_{v}$ are positive propagation constants, $\tau _{0}$ is
an initial temporal distance between the two solitons, and functions $U$ and
$V$, which may be assumed real, satisfy equations%
\begin{eqnarray}
k_{u}U+\frac{1}{2}\left( -\frac{d^{2}}{d\tau ^{2}}\right) ^{\alpha
/2}U+U^{3} &=&0,  \label{U} \\
k_{v}V+\frac{1}{2}\left( -\frac{d^{2}}{d\tilde{\tau}^{2}}\right) ^{\alpha
/2}V+V^{3} &=&0.  \label{V}
\end{eqnarray}%
In this work, we focus on the case of
\begin{equation}
k_{u}=k_{v}\equiv k=1,  \label{kkk}
\end{equation}%
when two equations (\ref{U}) and (\ref{V}) and the respective solitons are
actually identical, while $k=1$ may be fixed by scaling. Indeed, in WDM
systems solitons with equal energies are normally launched into different
channels \cite{WDM1}-\cite{Turitsyn}.

In the case of the ordinary (non-fractional) GVD, with $\alpha =2$, the
commonly known classical soliton solutions for $k_{u}=k_{v}=1$ are
\begin{equation}
U_{\mathrm{sol}}=\sqrt{2}\mathrm{sech}\left( \sqrt{2}\tau \right) ,V_{
\mathrm{sol}}=\sqrt{2}\mathrm{sech}\left( \sqrt{2}\tilde{\tau}\right) ,
\label{U,V}
\end{equation}%
the energy of each one being $E_{u}=E_{v}=2\sqrt{2}$. On the other hand, in
the same case of $\alpha =2$ but $c=0$, another obvious solution of Eqs. (%
\ref{U}) and (\ref{V}) is the two-component soliton,%
\begin{equation}
u=v=\sqrt{\frac{2}{3}}e^{iz}\mathrm{sech}\left( \sqrt{2}\tau \right) ,
\label{c=0}
\end{equation}%
with energies $E_{u}=E_{v}=2\sqrt{2}/3$.

\section{Soliton-soliton collisions}

For values of LI in interval (\ref{LI2}), single-component soliton solutions
to Eqs. (\ref{U}) and (\ref{V}) with $k_{u}=k_{v}=1$ were found by means of
the squared-operator iteration method \cite{Taras,Jianke}. These solutions
were used as the input for simulations of collisions between the $u$- and $v$%
-solitons, with a sufficiently large initial temporal separation $\tau _{0}$
between them, in the anticipation of the collision happening at $z\simeq
\tau _{0}/c$. The collisions were simulated by means of the split-step
algorithm, separating, at each step $\Delta z=10^{-5}\div 10^{-4}$ of
marching forward in $z$, increments to the solution generated by the linear
and nonlinear terms in Eqs. (\ref{2u}) and (\ref{2v}). To handle the linear
terms, the discrete Fourier transform was used for calculating $(-\partial
^{2}/\partial \tau ^{2})^{\alpha /2}$. Typically, the initial separation
between the solitons was taken as $\tau _{0}=8$, which is large enough in
comparison with the soliton's temporal width, that may be estimated as $%
W\sim (2k)^{-1/\alpha }$, pursuant to Eqs. (\ref{U}) and (\ref{V}) (hence,
for $k=1$ and $\alpha \leq 2$, $W$ takes values $\lesssim 1/\sqrt{2}\approx
0.7\allowbreak $). In most cases, the simulations were performed in the
domain of size $|\tau |\leq 128$, which was represented by a grid composed
of $4096$ points. This size was sufficient to unequivocally identify
outcomes of the collisions. The periodic boundary condition were used in the
simulations of the evolution. The accuracy of the simulations was confirmed
by monitoring the conservation of energies (\ref{EE}) and momentum (\ref{P})
(not shown in detail here).

\subsection{Collisions in the system with non-fractional GVD ($\protect%
\alpha =2$)}

\begin{figure}[h]
\centering\includegraphics[width=4.9in]{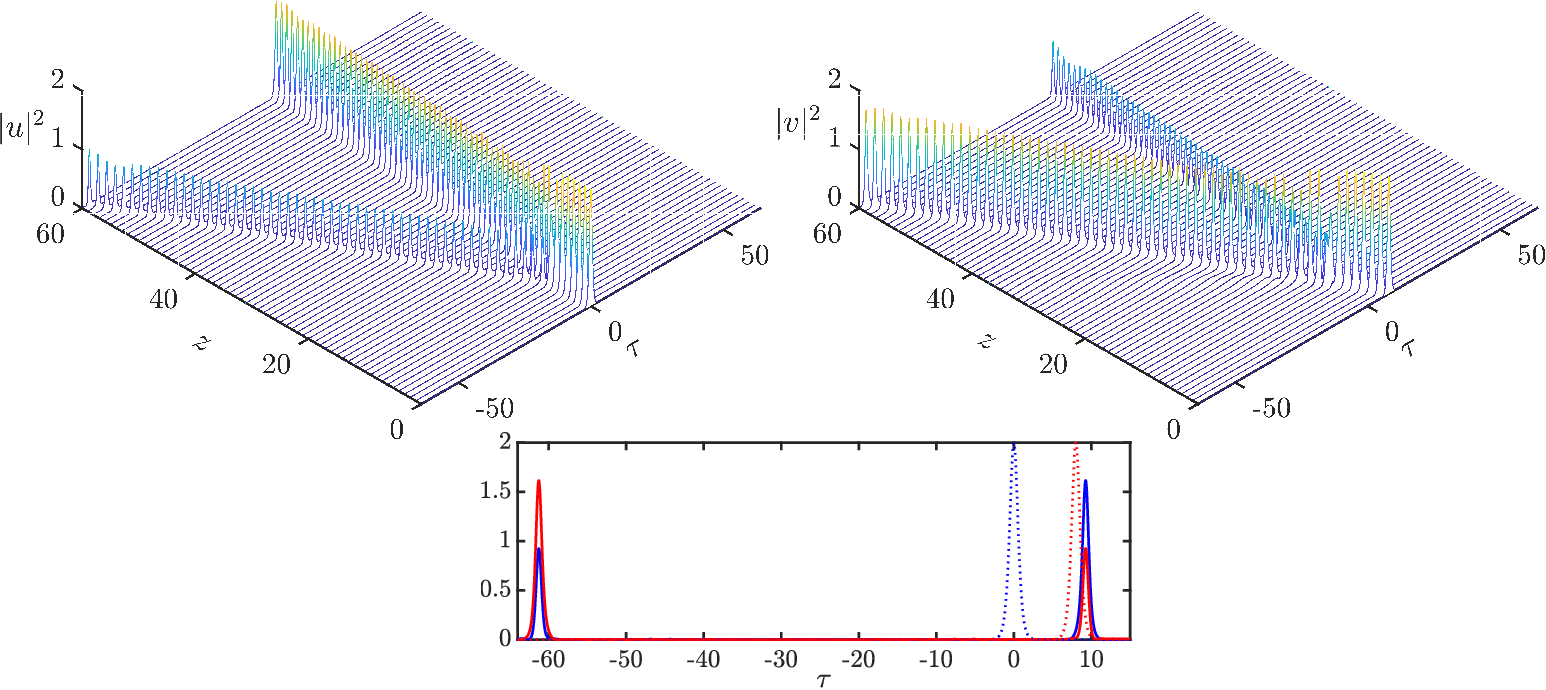}
\caption{The inelastic collision between the $u$- and $v$-component solitons
with the GV mismatch $c=-1,$ produced by simulations of Eqs. (\protect\ref%
{2u}) and (\protect\ref{2v}) with $\protect\alpha =2$ (the ordinary
non-fractional, GVD). The collision leads to the formation of a pair of
two-component solitons, which keep approximately the same rapidities, $0$
and $c$, as the single-component solitons had prior to the collision.}
\label{fig1}
\end{figure}

\begin{figure}[h]
\centering\includegraphics[width=4.9in]{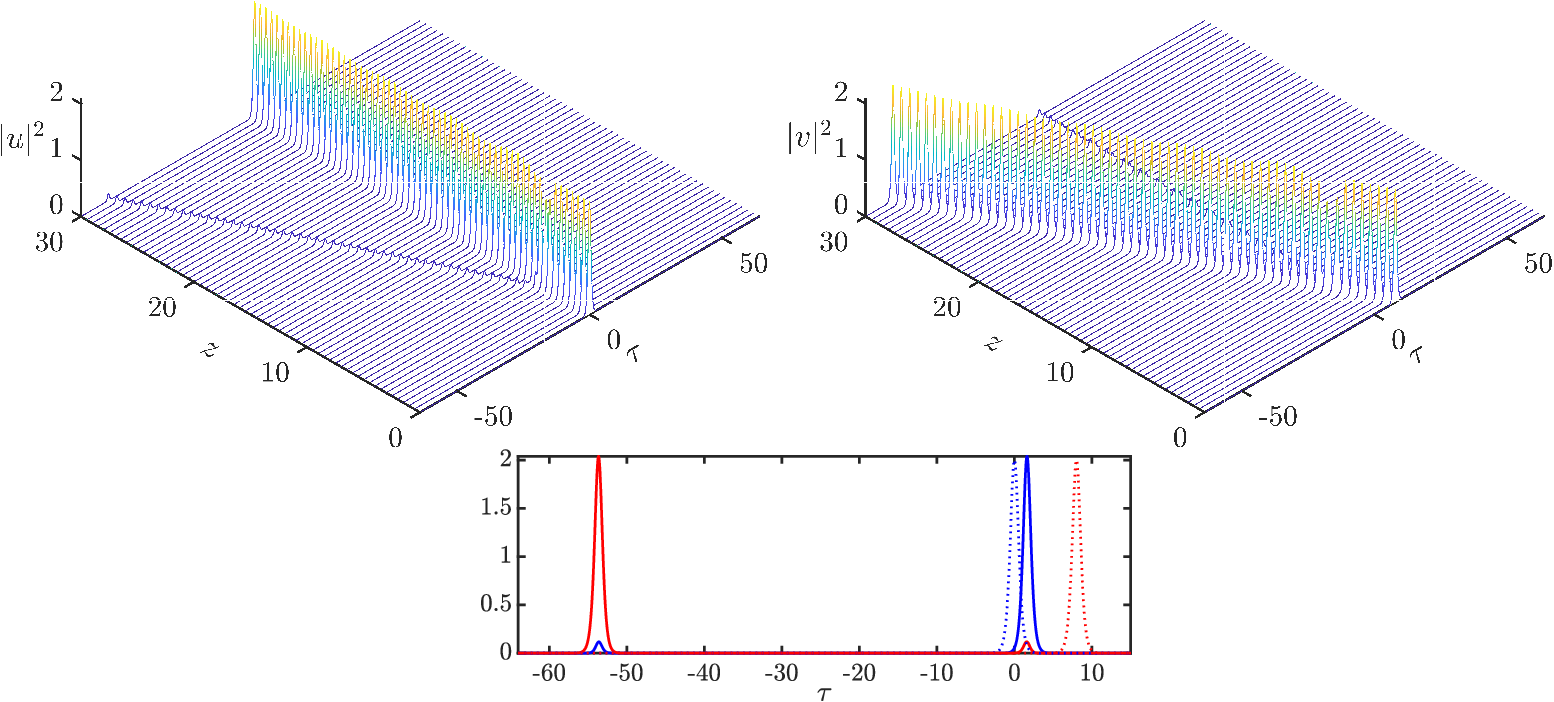}
\caption{The same as in Fig. \protect\ref{fig1}, but for $c=-2.0$. In this
case, the collision is nearly elastic.}
\label{fig2}
\end{figure}

First, for comparison with results produced by the FGVD system of Eqs. (\ref%
{2u}) and (\ref{2v}) with $\alpha <2$, we reproduce some known results for
soliton-soliton collisions in the system with the ordinary (non-fractional)
GVD, which corresponds to $\alpha =2$. Generally, collisions between
solitons in that system are inelastic for smaller values of $|c|$, and
quasi-elastic for larger $|c|$ \cite{Wabnitz}. The simulations of Eqs. (\ref%
{2u}) and (\ref{2v}) with the GV mismatch $|c|\lesssim 1$ demonstrate an
inelastic outcome in Fig. \ref{fig1} (for $c=-1$), \textit{viz}., the
formation of a pair of two-component solitons, with unequal energies of the
components. The emerging compound solitons keep approximately the same
rapidities ($0$ or $c$) which their dominant components had before the
collision. This outcome is also called ``splitting", with
the implication that each original single-component soliton \emph{splits} in
two fragments, which recombine into the pair of two-component solitons. At
larger $|c|$, Fig. \ref{fig2} demonstrates a nearly elastic collision for $%
c=-2$. In this case, the $u$- and $v$-solitons pass through each other, each
capturing a barely visible $v$- or $u$-components, respectively. As shown
below in Fig. \ref{fig13a}, in the case of $\alpha =2$ the collision becomes
completely elastic at $|c|\geq 2.6$.

\subsection{Collisions in the system with modern fractionality of the GVD, $%
\protect\alpha =1.5$}

\begin{figure}[h]
\centering\includegraphics[width=4.9in]{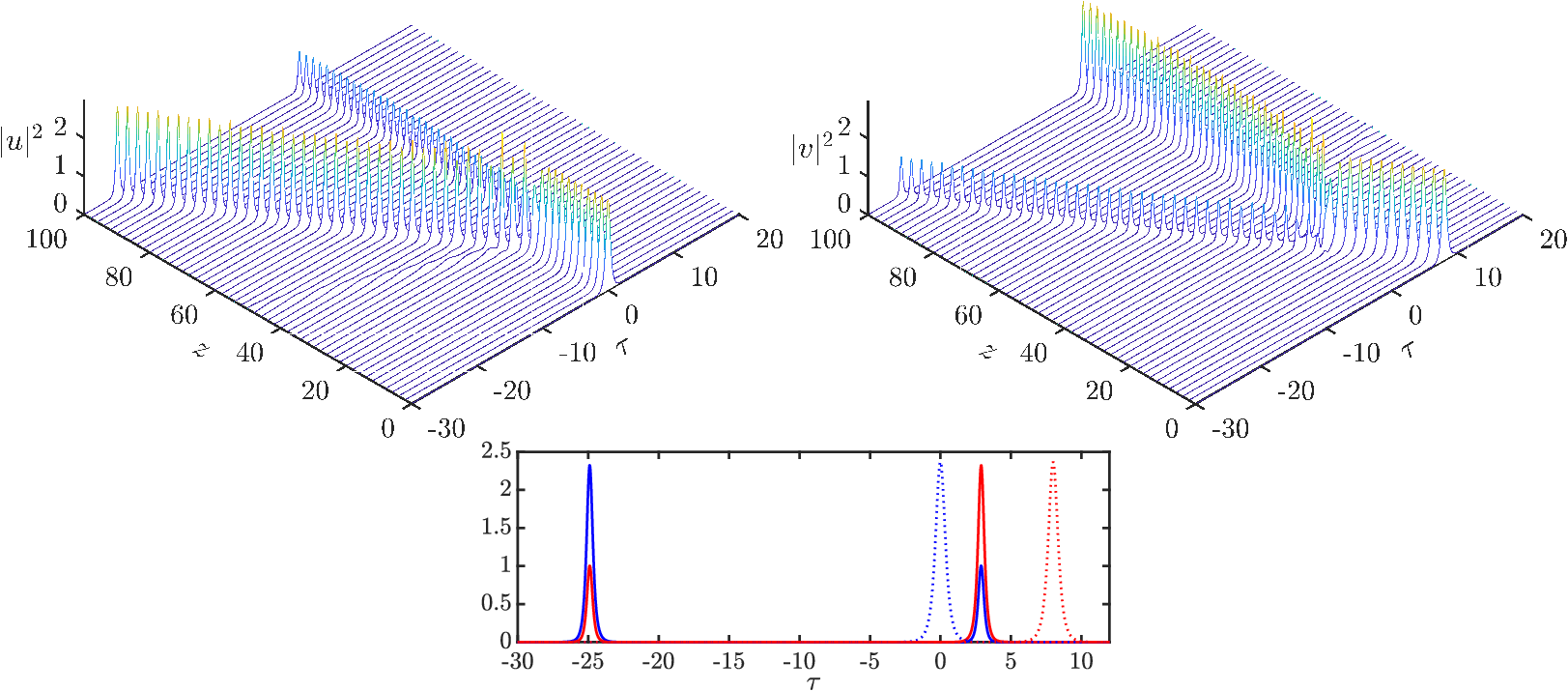}
\caption{Top panels: the collision between the $u$- and $v$-solitons with $%
c=-0.3$, as produced by the simulations of Eqs. (\protect\ref{2u}) and (%
\protect\ref{2v}) with $\protect\alpha =1.5$. The bottom panel: the soliton
profiles, $|u(\protect\tau )|^{2}$ and $|v(\protect\tau )|^{2}$, at the
input, $z=0$ (dotted lines), and in the final confuguration (solid lines).
This collision can be categorized as ``splitting", similar
to the one displayed, for $\protect\alpha =2$, in Fig. \protect\ref{fig1}.}
\label{fig3a}
\end{figure}

In the case of a moderate degree of the fractionality, corresponding,
typically, to the LI value of $\alpha =1.5$, the collisions with relatively
small values of $|c|$, such as $c=-0.3$, give rise to the outcome identified
above as ``splitting", i.e., conversion of the colliding
single-component solitons into a pair of compounds with unequal components,
as shown in Fig. \ref{fig3a}. Also similar to the above-mentioned example,
the rapidities of the emerging compound solitons remain very close to those
of their dominant ``parent" components, i.e.,  $0$ and $c$,
for the $u$- and $v$-components, respectively. In addition, the bottom panel
in Fig. \ref{fig3a} demonstrates that the collision gives rise to a shift of
each soliton as a whole, which is a well-known feature of soliton-soliton
interactions \cite{Agrawal,Jianke}. For instance, the center of the soliton
with zero rapidity shifts from the initial position, $\tau _{\mathrm{initial}%
}=0$, to the final one,  $\tau _{\mathrm{final}}\simeq 3$.

\begin{figure}[h]
\centering\includegraphics[width=4.9in]{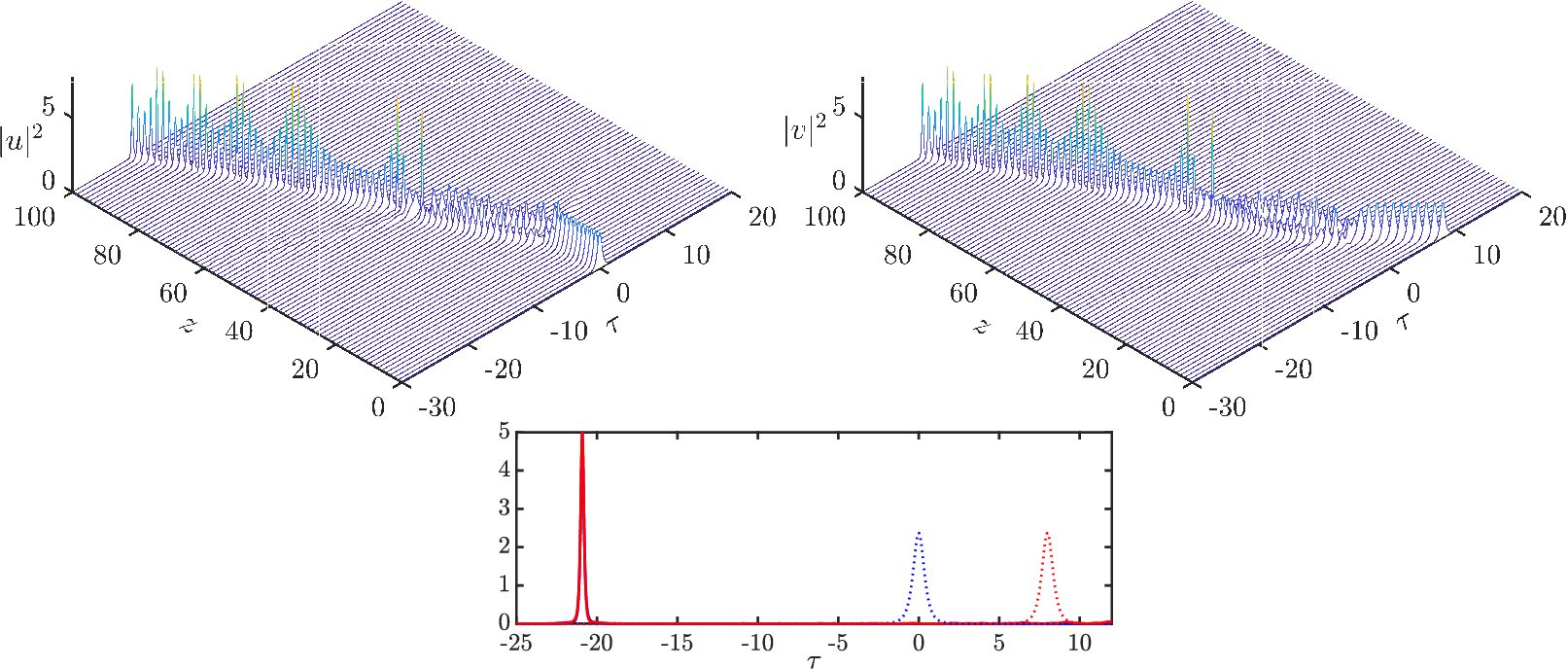}
\caption{Top panels: the merger of the colliding solitons into a breather
with strong internal vibrations, in the case of $\protect\alpha =1.5$ and $%
c=-0.5$. The bottom panel: the soliton profiles, $|u(\protect\tau )|^{2}$
and $|v(\protect\tau )|^{2}$, at the input, $z=0$ (dotted lines), and in the
final configuration (solid lines, which completely overlap for both
components). }
\label{fig4a}
\end{figure}

The increase of the rapidity from $c=-0.3$, corresponding to Fig. \ref{fig3a}%
, to $c=-0.5$ leads to the \textit{complete merger} of the colliding
solitons into a bound state, as shown in Fig. \ref{fig4a}. It is seen that
the merger proceeds via the fission and final fusion of the emerging bound
state. It appears in the form of a \textit{breather}, with strong internal
vibrations. In agreement with the momentum conservation (see Eq. (\ref{P})),
the bound state moves with rapidity $\simeq c/2$, which is the mean value of
the rapidities of the original solitons which merge into the bound state.

The merger of the colliding solitons takes place in a narrow interval of
values of the GV mismatch $c$ (see Fig. \ref{fig13a} below). As shown in
Fig. \ref{fig5a}, the increase of $|c|$ from $0.5$, which corresponds to
Fig. \ref{fig4a}, to $c=-0.6$ switches the outcome into the
``splitting" regime, which is relatively close to the
elastic one, i.e., featuring a pair of compound solitons with a small $v$%
-component attached to the large $u$-component, and vice versa, the
rapidities staying close to the ``parent" values, i.e., $0$
and $c$, respectively (cf. the outcome displayed in Fig. \ref{fig2} for $%
\alpha =2$ and $c=-1$). This collision regime is observed, for $\alpha =1.5$%
, in two intervals of the rapidity, \textit{viz}., $-0.9\leq c\leq -0.6$ and
$-1.8\leq c\leq -1.1$. In the narrow gap between them, the ``splitting" outcome is different. As shown in Fig. \ref{fig6a} for $c=-1.0$,
this case gives rise to the creation of a pair of compound solitons with
nearly equal energies of both components. The rapidities of the compound
states are seen to be very close to the initial values, i.e., $0$ for one
compound, and $c$ for the other. Finally, in the case of $\alpha =1.5$ the
collisions are completely elastic at $c>|1.8|$.

\begin{figure}[h]
\centering\includegraphics[width=4.9in]{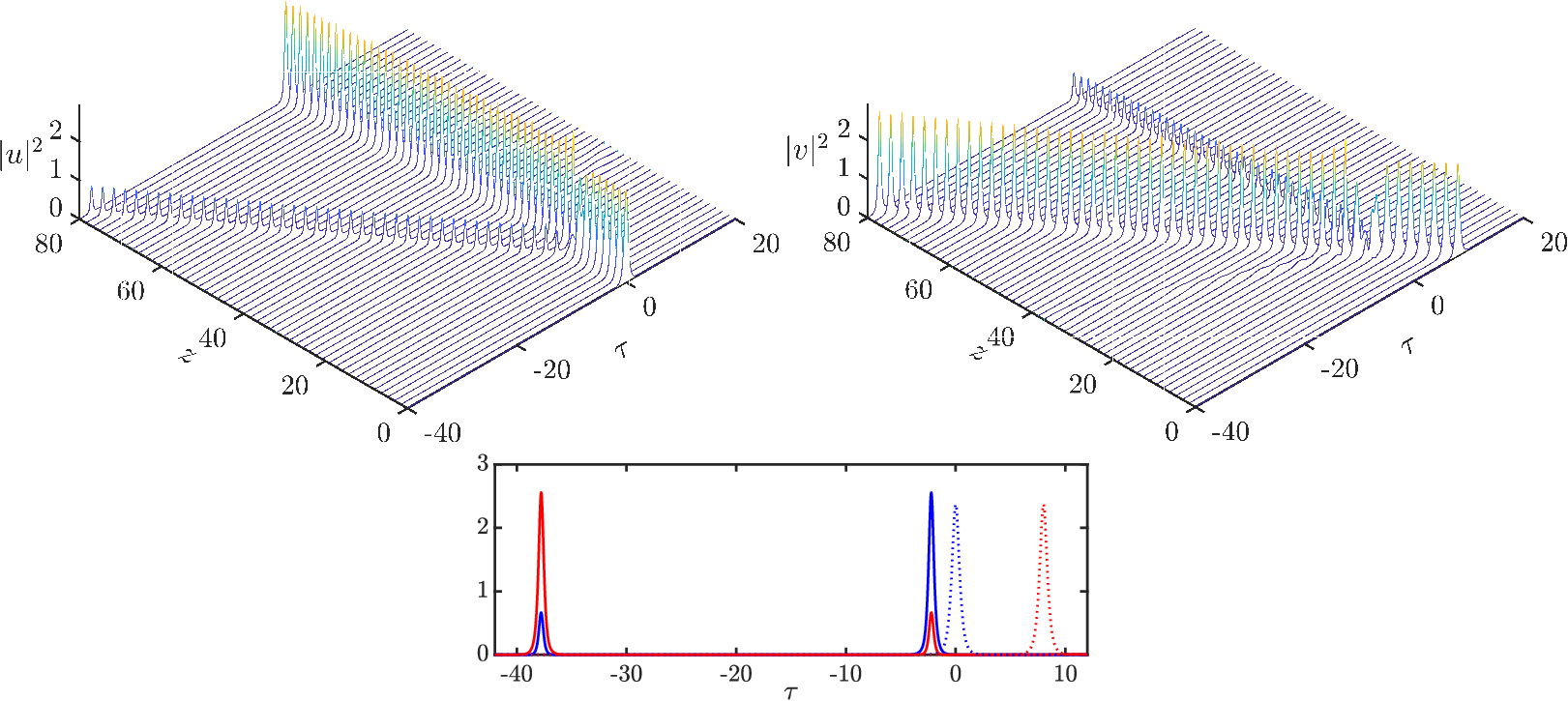}
\caption{Top panels: the nearly elastic ``splitting" of the
colliding solitons, i.e., the generation of two-component solitons, with a
small $v$ -component in one, and a small $u$-component in the other, in the
case of $\protect\alpha =1.5$ and $c=-0.6$. The bottom panel: the soliton
profiles, $|u(\protect\tau )|^{2}$ and $|v(\protect\tau )|^{2}$, at the
input, $z=0$ (dotted lines), and in the final confuguration (solid lines).}
\label{fig5a}
\end{figure}

\begin{figure}[h]
\centering\includegraphics[width=4.9in]{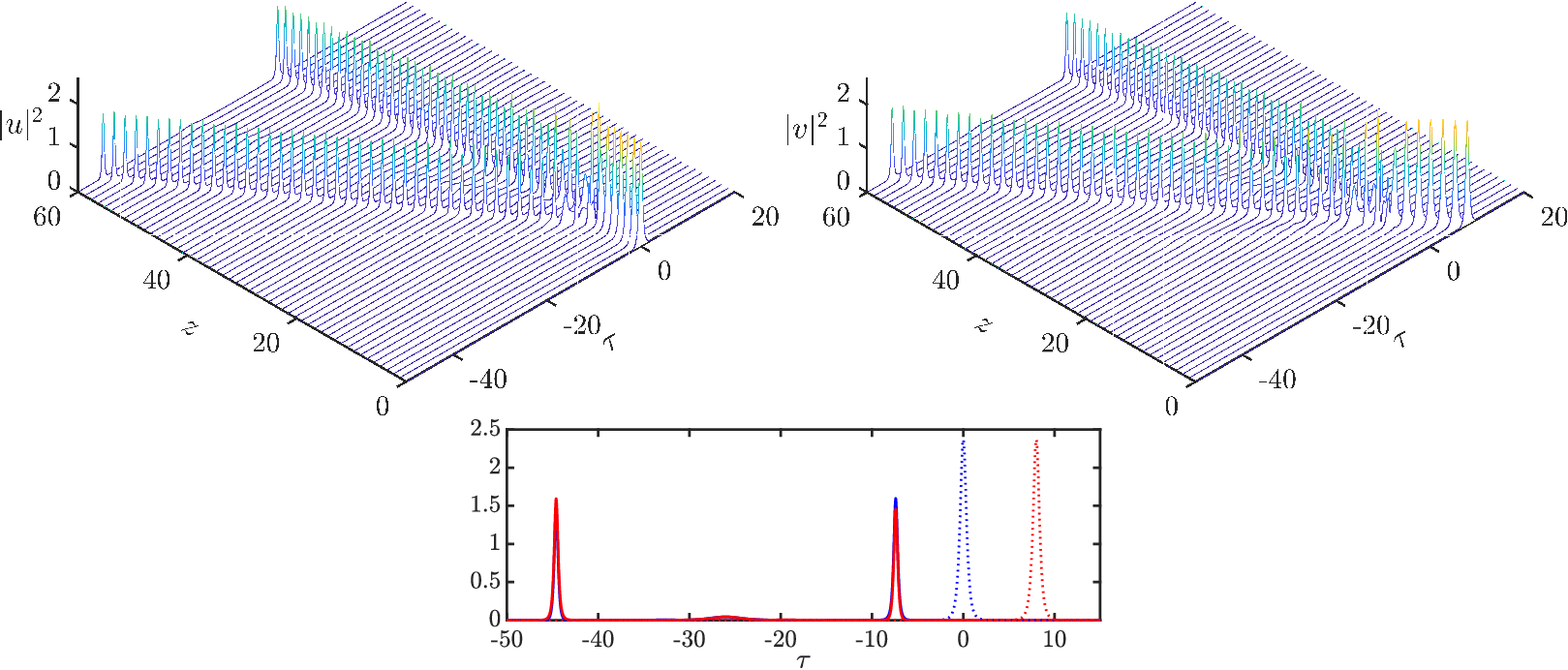}
\caption{Top panels: the special case of the ``splitting",
namely, the transformation of the colliding solitons into a pair of compound
ones, each with nearly equal energies of both components, in the case of $%
\protect\alpha =1.5$ and $c=-1.0$. The bottom panel: the soliton profiles, $%
|u(\protect\tau )|^{2}$ and $|v(\protect\tau )|^{2}$, at the input, $z=0$
(dotted lines), and in the final confuguration (solid lines, which
completely overlap for both components in each compound soliton).}
\label{fig6a}
\end{figure}

\subsection{Collisions in the systems with strong fractionality, $\protect%
\alpha =1.2$ and $1.1$}

\begin{figure}[h]
\centering\includegraphics[width=4.9in]{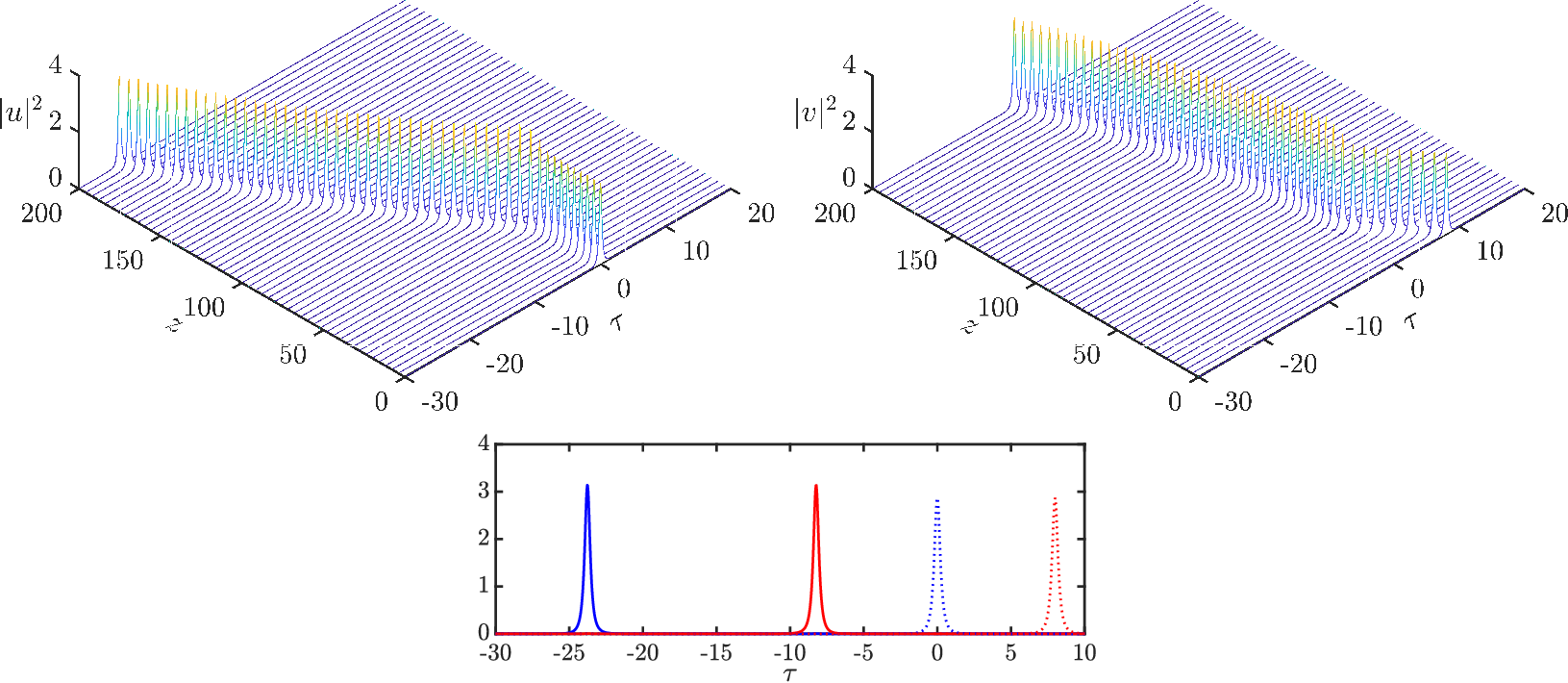}
\caption{Top panels: the rebound of the colliding solitons in the case of $%
\protect\alpha =1.2$ and $c=-0.2$. The bottom panel: the soliton profiles, $%
|u(\protect\tau )|^{2}$ and $|v(\protect\tau )|^{2}$, at the input, $z=0$
(dotted lines), and in the final confuguration (solid lines).}
\label{fig7a}
\end{figure}

The system with stronger fractionality -- in particular, with $\alpha =1.2,$
which is relatively close to the left edge of interval (\ref{LI2}) of the
admitted values of LI, features different outcomes of the collisions. First,
at relatively small values of $|c|$ the collision seems elastic, but not in
the same sense as above: as shown in Fig. \ref{fig7a} for $c=-0.2$, the
solitons interact with strong repulsion, bouncing back from each other,
similar to the collision of hard mechanical particles. Thus, the solitons
approximately exchange their velocities, so that the $u$-soliton, which
originally had rapidity $0$, is set in motion with rapidity close to but
smaller than $c$, while the $v$-soliton features a strong drop of the
rapidity from the initial value, $c$. This outcome occurs in the interval of
$|c|<0.3$.

\begin{figure}[h]
\centering\includegraphics[width=4.9in]{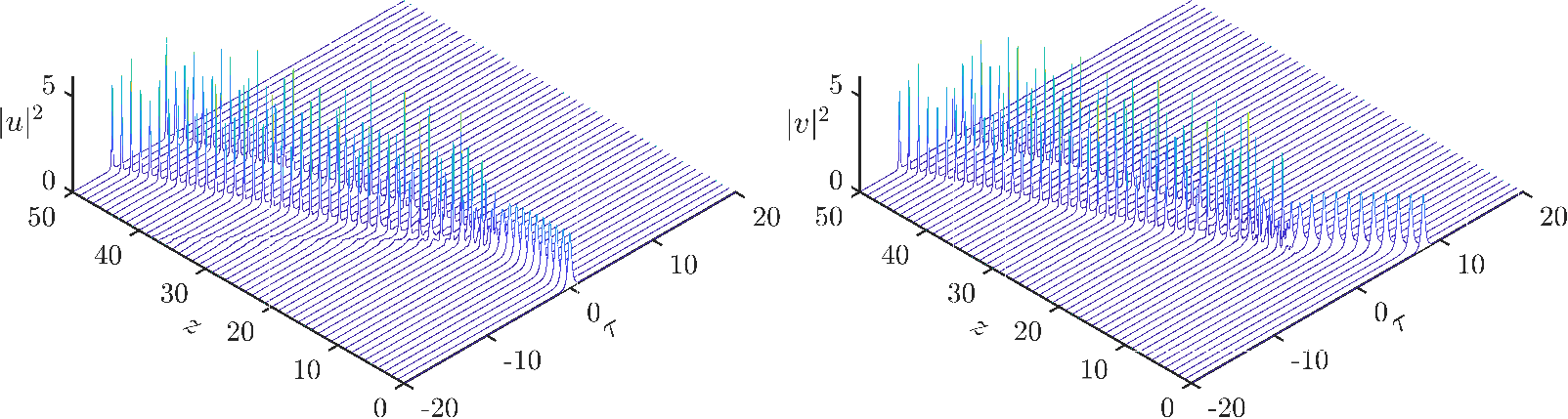}
\caption{The ``splitting" of the colliding solitons into
two-component breathers with close values of the rapidity (intermediate
between the initial ones, $0$ and $c$), in the case of $\protect\alpha =1.2$
and $c=-0.7$.}
\label{fig8a}
\end{figure}

In the region of $|0.3|\leq c\leq |1.2|$, the collision leads to the outcome
which was defined above as ``splitting", i.e., formation of
a pair of compound solitons with unequal energies of the components. As
shown in Fig. \ref{fig8a} for $c=-0.7$, the difference from the
``splitting" outcomes displayed above (see Figs. \ref{fig1}, %
\ref{fig3a}, \ref{fig5a}, and \ref{fig6a}) is that the two-component
solitons are generated in the form of \textit{breathers}, featuring strong
internal vibrations. The rapidities of the two breathers take close values,
both being intermediate between the initial ones, $0$ and $c$.

\begin{figure}[h]
\centering\includegraphics[width=4.9in]{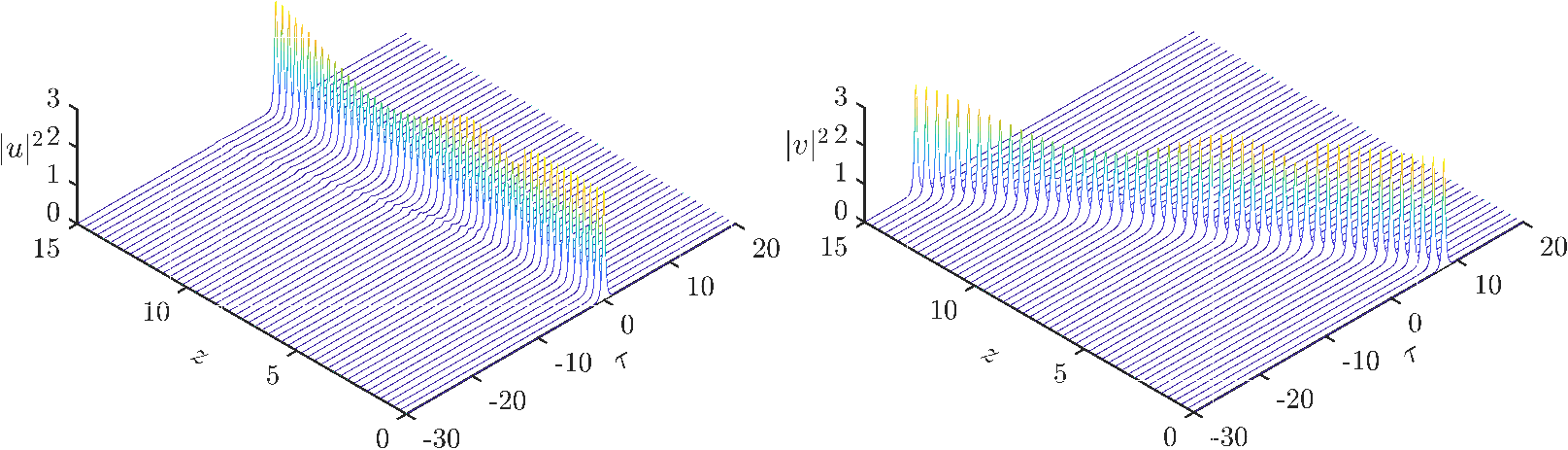}
\caption{The quasi-elastic collision of solitons, leading to excitation of
low-frequency intrinsic oscillations in them, in the case of $\protect\alpha %
=1.2$ and $c=-2.0$.}
\label{fig9a}
\end{figure}

In the adjacent interval, $|1.3|<c\leq |3.0|$, Fig. \ref{fig9a} demonstrates
(for $c=-2.0$) that the collisions are, generally speaking, quasi-elastic,
but with a peculiarity: after the mutual passage, the solitons develop
intrinsic low-amplitude excitations. Finally, at $|c|>3$, the simulations of
Eqs. (\ref{2u}) and (\ref{2v}) with $\alpha =1.2$ produce completely elastic
collisions.

\begin{figure}[h]
\centering\includegraphics[width=4.9in]{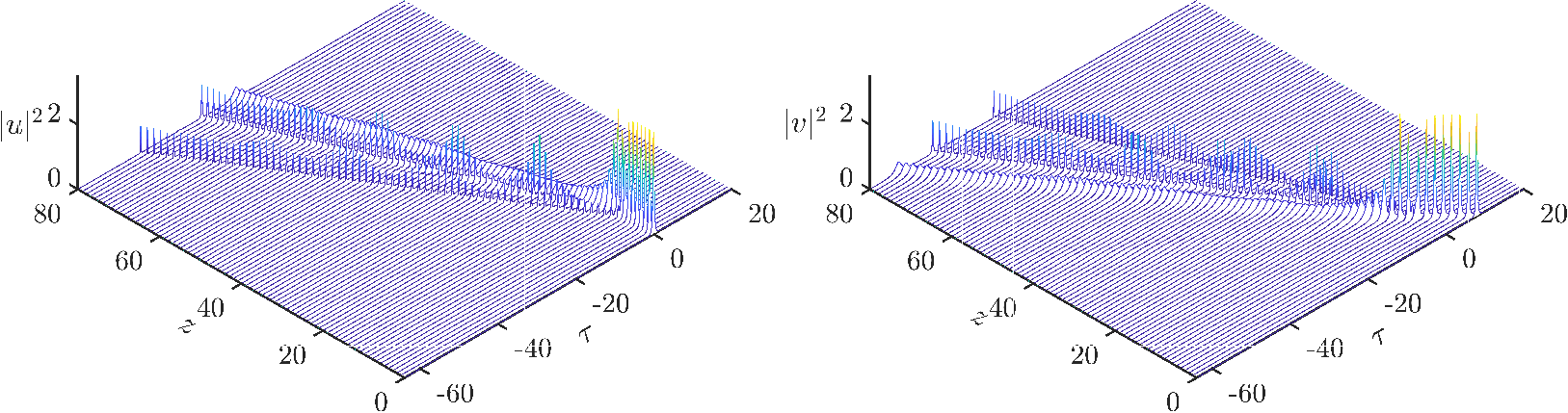}
\caption{The collision of solitons in the case of $\protect\alpha =1.1$ and $%
c=-1.1$, leading to their ``splitting" into a set of three
breathers with different rapidities.}
\label{fig10a}
\end{figure}

\begin{figure}[h]
\centering\includegraphics[width=4.9in]{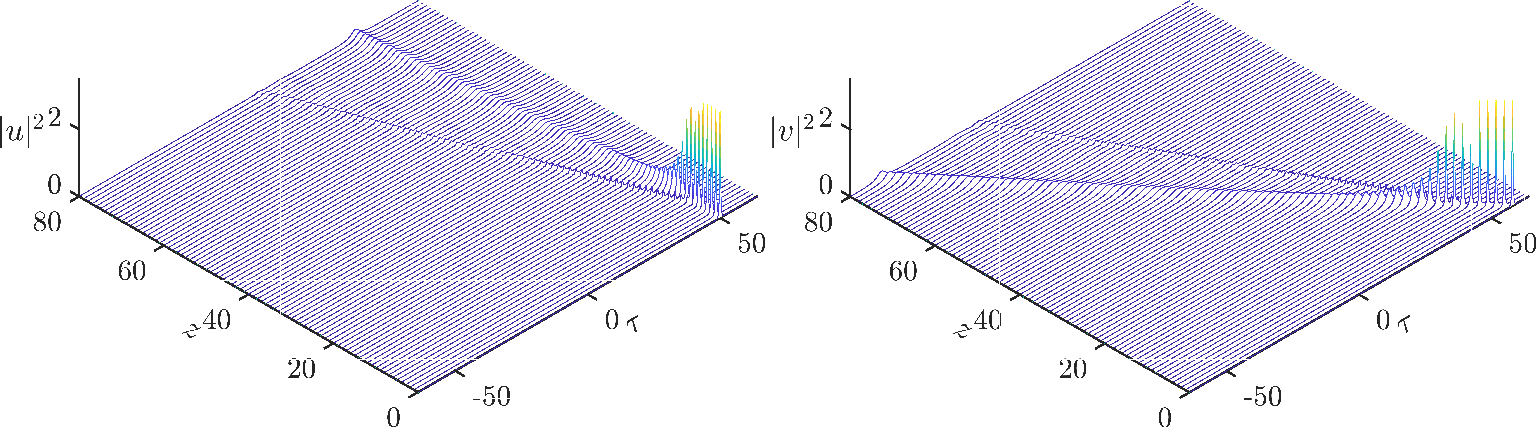}
\caption{The collision of solitons in the case of $\protect\alpha =1.1$ and $%
c=-1.5$, leading to the extreme form of ``splitting", close
to full destruction of the colliding solitons.}
\label{fig11a}
\end{figure}

\begin{figure}[h]
\centering\includegraphics[width=4.9in]{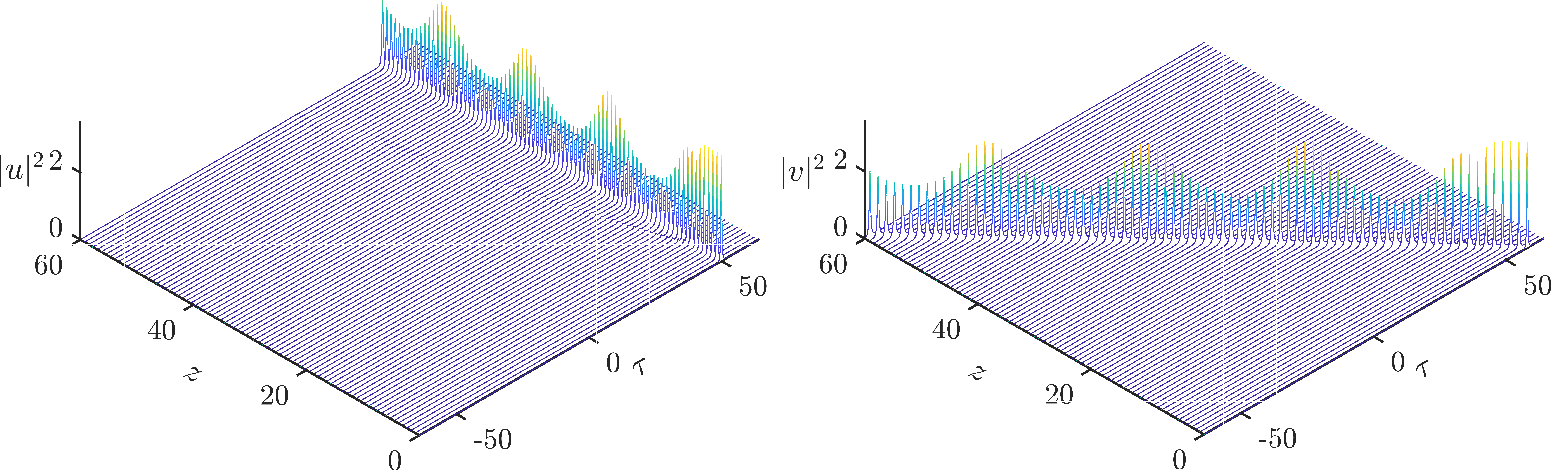}
\caption{The quasi-elastic collision of solitons in the case of $\protect%
\alpha =1.1$ and $c=-2.0$. The solitons reappear in excited states,
featuring low-frequency intrinsic vibrations, with approximately equal
frequencies and amplitudes.}
\label{fig12a}
\end{figure}

The decrease of LI from $\alpha =1.2$ to $\alpha =1.1$, i.e., moving
essentially closer to the left edge of interval (\ref{LI2}), leads to
drastic changes in outcomes of the collisions. In particular, the
``splitting" now produces not two, but three solitons, all featuring strong
intrinsic vibrations (breathers), as shown in Fig. \ref{fig10a} for $c=-1.1$%
. With the further increase of the rapidity, the ``splitting" degenerates
into nearly full destruction of the colliding solitons, see Fig. \ref{fig11a}
for $c=-1.5$. In the case of $\alpha =1.1$, quasi-elastic collisions take
place at $|c|>$ $1.6$, featuring the excitation of intrinsic low-frequency
vibrations in the solitons, as shown for $c=-2$ in Fig. \ref{fig12a} (cf.
Fig. \ref{fig9a}).

\subsection{The summary of the results for the soliton-soliton collisions}

\begin{figure}[h]
\centering\includegraphics[width=2.8in]{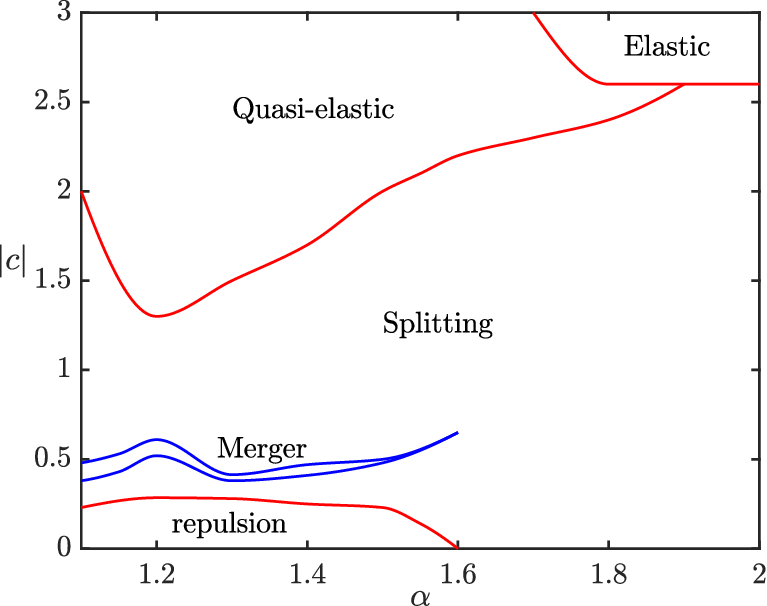}
\caption{The chart of outcomes of the soliton-soliton collisions in the
plane of LI, $\protect\alpha $, and GV mismatch, $c$, produced by systematic
simulations of Eqs. (\protect\ref{2u}) and (\protect\ref{2v}).}
\label{fig13a}
\end{figure}

Outcomes of the collision, produced by systematic simulations of the
underlying system of XPM-coupled FNLSEs (\ref{2u}) and (\ref{2v}), are
summarized in the parameter plane of LI and GV$\ $mismatch, $(\alpha ,c)$,
which is displayed in Fig. \ref{fig13a}. In this chart, the
``repulsion" outcome, located near the left bottom edge of
the parameter plane, implies the rebound of the colliding solitons, see the
example in Fig. \ref{fig7a}. The ``merger", which takes part
in the narrow strip at $\alpha \leq 1.6$ and $|c|$ close to $0.5$, is
exemplified by the picture shown in \ Fig. \ref{fig4a}, in which the
colliding solitons merge into a single breather.

The vast area of ``splitting" in Fig. \ref{fig13a} includes the cases when
the collision converts the single-component solitons into a pair of
two-component ones, such as those displayed in Figs. \ref{fig1}, \ref{fig3a}%
, and \ref{fig6a}. Also belonging to this area are outcomes in which the
emerging two-component solitons are close to the single-component ones, each
featuring one large and one small components (Figs. \ref{fig2} and \ref%
{fig5a}), and, on the other hand, the cases in which two-component solitons
appear in the form of breathers with strong intrinsic vibrations, as in Fig. %
\ref{fig8a}. In addition, the ``splitting area" includes a small sub-area
(found only for $\alpha =1.1$, i.e., in the case of the strongest
fractionality) in which the colliding solitons split in a set of three
breathers (Fig. \ref{fig10a}) or suffer nearly complete destruction, as in
Fig. \ref{fig11a}.

The ``quasi-elastic" region in Fig. \ref{fig13a} includes
the cases when the colliding solitons pass through each other, reappearing
in the state with low-frequency intrinsic vibrations, such as in Figs. \ref%
{fig9a} and \ref{fig12a}. Examples of fully elastic collisions, which occur
at large values of rapidity $|c|$ and/or for LI close to the ordinary
(non-fractional) value $\alpha =2$ (the top right corner in Fig. \ref{fig13a}%
) are not displayed here, as the picture is obvious: the two solitons merely
pass through each other, featuring a small shift of their centers, but no
changes in their structure or rapidities.

\section{Two-component bound states}

The numerical results presented above in Figs. \ref{fig1}, \ref{fig3a}, and %
\ref{fig6a}, as well as in Figs. \ref{fig2} and \ref{fig5a}, demonstrate the
formation of two-component (compound) solitons, in which the constituents
build a robust bound state, in spite of the action of the GV mismatch $c$,
which tends to split the bound state. In this section, we aim to produce
families of such compound solitons directly (not as ``accidental" results of the collisions between the single-component
solitons), and systematically explore their properties.

Numerical solutions for the bound states were produced by means of a
two-stage procedure. First, ansatz
\begin{equation}
u=\exp \left( ikz\right) U(\tau ),~v=\exp \left( ikz\right) V(\tau ),
\label{uv}
\end{equation}%
with $k=1$ (as per Eq. (\ref{kkk})), is substituted in the system of
XPM-coupled FNLSEs (\ref{2u}) and (\ref{2v}), leading to ordinary
differential equations%
\begin{gather}
U+\frac{1}{2}\left( -\frac{d^{2}}{d\tau ^{2}}\right) ^{\alpha /2}U+\left(
|U|^{2}+2|V|^{2}\right) U=0,  \label{UV1} \\
V-ic\frac{dV}{d\tau }+\frac{1}{2}\left( -\frac{d^{2}}{d\tau ^{2}}\right)
^{\alpha /2}V+\left( |V|^{2}+2|U|^{2}\right) V=0,  \label{UV2}
\end{gather}%
cf. Eqs. (\ref{U}) and (\ref{V}). In the absence of the XPM coupling, Eqs. (%
\ref{UV1}) and (\ref{UV2}) would produce, severally, single-component
solitons such that, if used as initial conditions for the uncoupled
equations (\ref{2u}) and (\ref{2v}), would give rise to a quiescent $u$%
-soliton and the $v$-soliton moving with rapidity $c$.

\begin{figure}[h]
\centering\includegraphics[width=4.9in]{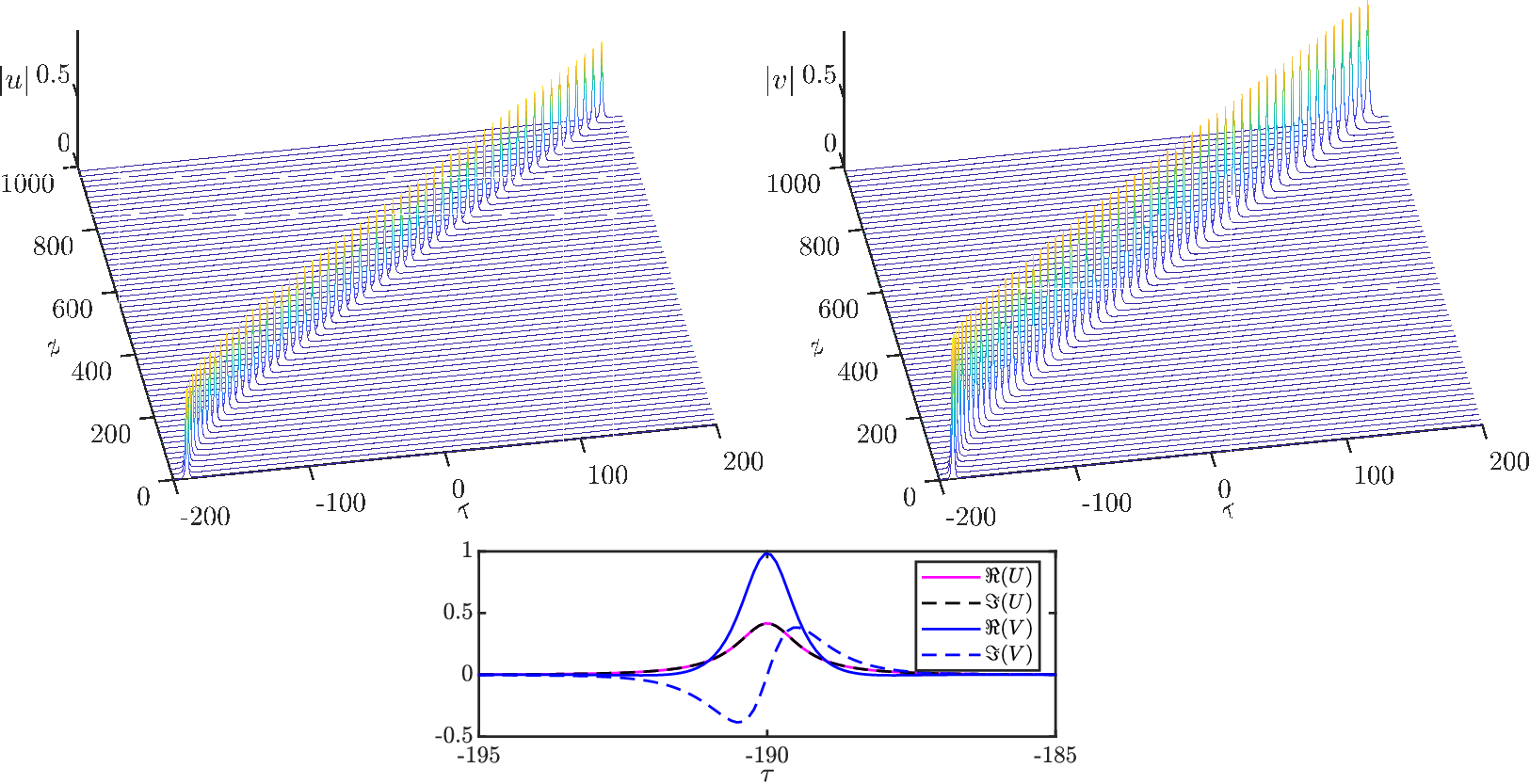}
\caption{Left panels: the relaxation of the two-component soliton towards
the stationary form, as produced by the numerical simulation of Eqs. (
\protect\ref{2u}) and (\protect\ref{2v}), starting from the input produced
as a numerical solutions of Eqs. (\protect\ref{U}) and (\protect\ref{V}),
with $\protect\alpha =1.5$ and $c=0.7$. The real and imaginary parts of $U(%
\protect\tau )$ and $V(\protect\tau ))$, used as the input, are displayed in
the right panel.}
\label{fig14a}
\end{figure}

At the second stage, the two-component solution of Eqs. (\ref{UV1}) and (\ref%
{UV2}) was used as the input for the full (XPM-coupled) system of FNLSEs$.$
As a result, the two-component bound state moves with deceleration and
slightly varying amplitudes, quickly relaxing towards a stable stationary
two-component soliton, with a constant rapidity and constant amplitudes of
its components. A typical example of the dynamical relaxation is shown in
Fig. \ref{fig14a}, for $\alpha =1.5$ and $c=0.7$.

\begin{figure}[h]
\centering\includegraphics[width=4in]{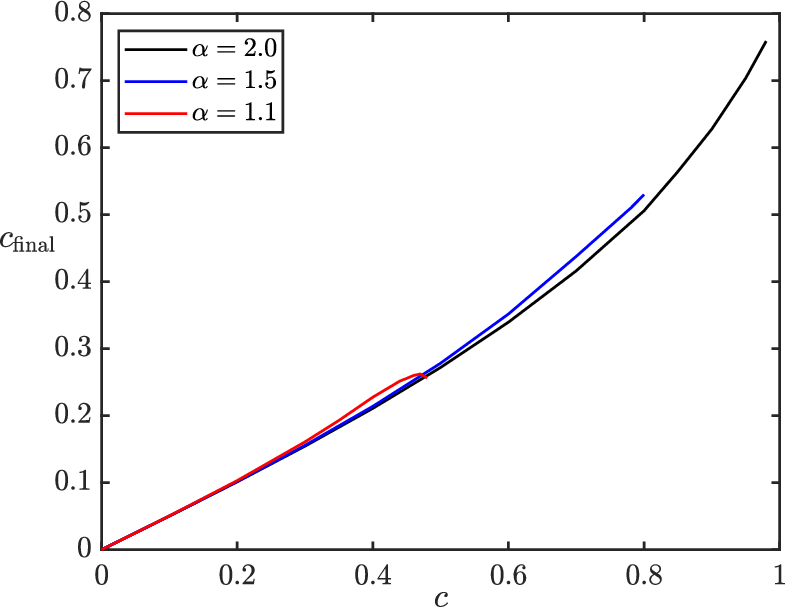}
\caption{The final rapidity of the established two-component soliton, $c_{%
\mathrm{final}}$, vs. the GV mismatch parameter $c$ in the underlying system
of Eqs. (\protect\ref{2u}) and (\protect\ref{2v}), for three values of LI, $%
\protect\alpha =2$ (the ordinary non-fractional GVD), $\protect\alpha =1.5$
(modern fractionality), and $\protect\alpha =1.1$ (strong fractionality).}
\label{fig15a}
\end{figure}

\begin{figure}[h]
\centering\includegraphics[width=4.9in]{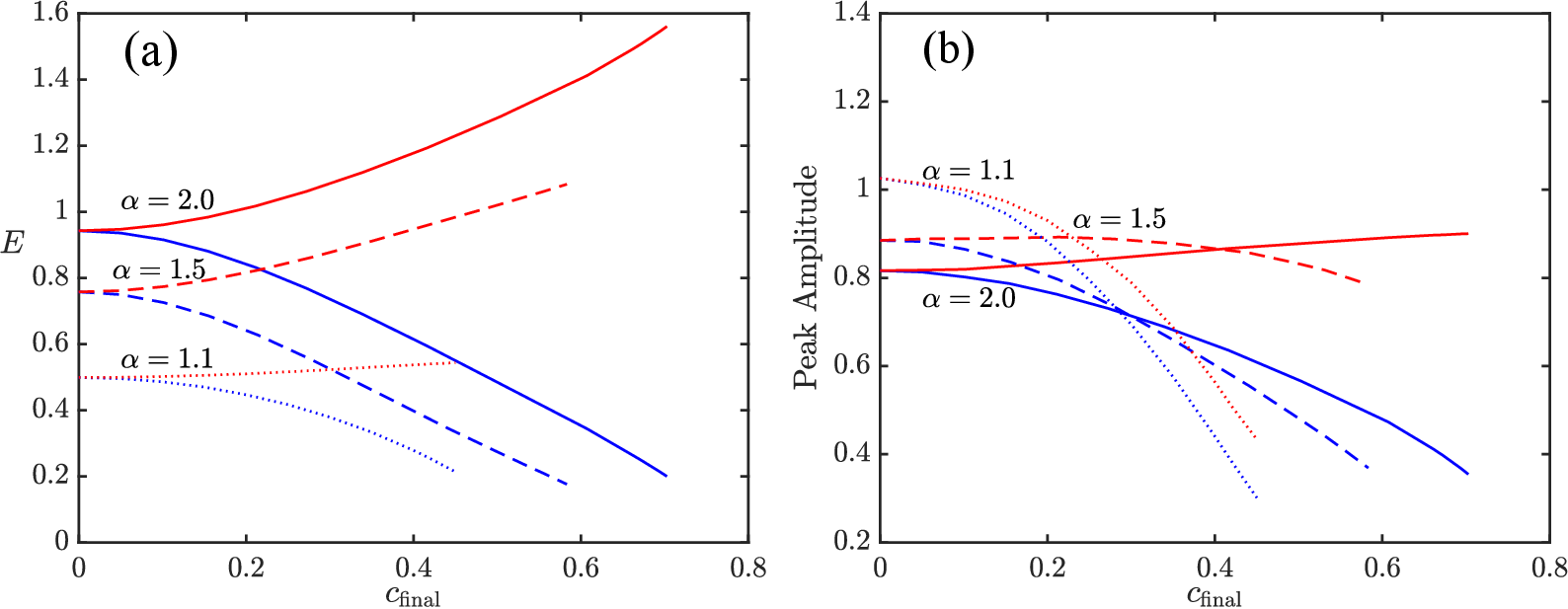}
\caption{The energies (a) and amplitudes (b) of the established $u$- and $v$%
- components (blue and red curves, respectively) of the compound solitons
vs. the their established rapidity, $c_{\mathrm{final}}$, for the same LI
values as in Fig. \protect\ref{fig15a}.}
\label{fig16a}
\end{figure}

The results of the numerical analysis are summarized in Fig. \ref{fig15a},
by dint of the dependence of the final rapidity, $c_{\mathrm{final}}$, of
the established two-component soliton, on the GV mismatch $c$ in Eqs. (\ref%
{2u}) and (\ref{2v}), for three different values of LI, \textit{viz}., $%
\alpha =2$, $1.5$, and $1.1$. Naturally, $c_{\mathrm{final}}$ is smaller
than $c$, as it is a result of the interplay of the $u$- and $v$-components,
whose proper values of the rapidity are $0$ and $c$, respectively. For the
same three values of LI, additional summary diagrams are provided by Fig. %
\ref{fig16a}, in which the energies and amplitudes of both components of the
established two-component solitons are plotted vs. $c_{\mathrm{final}}$.

The curves plotted in Figs. \ref{fig15a} and \ref{fig16a} terminate at
points where the amplitude of the $u$-component becomes too small. The
interval of the rapidities in which the two-component solitons exist shrinks
with the decrease of LI, in agreement with the fact that the GVD
fractionality destroys the Galilean invariance of Eqs. (\ref{2u}) and (\ref%
{2v}), thus impeding the creation of moving solitons \cite{review,review2}.

\section{Conclusion}

We\ have considered the copropagation of optical pulses, carried by
different wavelengths or orthogonal circular polarizations of light, in the
fiber cavity with the effective FGVD (fractional group-velocity dispersion),
which is currently available to experiments. The system is modeled by the
XPM (cross-phase-modulation)-coupled system of FNLSEs (fractional nonlinear
Schr\"{o}dinger equations), with the FGVD represented by the Riesz
fractional derivatives, whose LI (L\'{e}vy index), $\alpha $, takes values $%
1<\alpha \leq 2$ (the ordinary non-fractional GVD corresponds to $\alpha =2$%
). The system includes the SPM\ (self-phase-modulation) nonlinear terms and
the GV (group-velocity) mismatch $c$. By means of robust numerical methods,
we have identified areas in the plane of $\left( \alpha ,c\right) $ for
different outcomes of the soliton-soliton collisions: mutual rebound,
``splitting", i.e., the transformation of the colliding
single-component solitons into a pair of two-component ones, merger of the
solitons into a breather, quasi-elastic passage of the solitons with
excitations of intrinsic vibrations in both of them, and fully elastic
collisions. Using another numerical procedure, families of two-component
solitons are constructed systematically, identifying their rapidity and
energies of their components.

As an extension of the analysis, it may be interesting to study in detail
internal modes of the single- and two-component solitons, the excitation of
which is demonstrated by numerical results reported in this work.

\section{Acknowledgments}

T.Z. would like to express her gratitude to His Majesty the King of Bhutan
and Naresuan University for granting the fully funded Master's Degree
scholarship. The Work of T.M. is supported by Faculty of Engineering,
Naresuan University (Grant number R2567E039). The work of B.A.M. is
supported, in part, by the Israel Science Foundation through grant No.
1695/22.

\end{document}